\documentclass[aps,prb,twocolumn,superscriptaddress]{revtex4-2}
\usepackage{graphicx}
\usepackage{hyperref}
\usepackage{array}
\usepackage{hhline}
\usepackage{dcolumn}
\usepackage{bm}
\usepackage{braket}
\usepackage{amsmath}
\usepackage{amssymb}
\usepackage[version=4]{mhchem}

\hypersetup{
colorlinks = true,
urlcolor = blue,
linkcolor = blue,
citecolor = blue
}
\begin{document}
\title{
  Nonlinear phononics in LaFeAsO: Optical control of the crystal structure toward possible enhancement of superconductivity
}
\author{Shu Kamiyama}
\email{kamiyama@presto.phys.sci.osaka-u.ac.jp}
\affiliation{Department of Physics, The University of Osaka, Toyonaka, Osaka 560-0043, Japan}
\author{Tatsuya Kaneko}
\affiliation{Department of Physics, The University of Osaka, Toyonaka, Osaka 560-0043, Japan}
\author{Kazuhiko Kuroki}
\affiliation{Department of Physics, The University of Osaka, Toyonaka, Osaka 560-0043, Japan}
\author{Masayuki Ochi}
\affiliation{Department of Physics, The University of Osaka, Toyonaka, Osaka 560-0043, Japan}
\affiliation{Forefront Research Center, The University of Osaka, Toyonaka, Osaka 560-0043, Japan}
\date{\today}

\begin{abstract}
  Nonlinear phononics provides a route to control crystal structures through light-induced phonon excitation.
  In this study, we apply nonlinear phononics to an iron-based superconductor, LaFeAsO, with the aim of tuning its crystal structure toward the ideal one to enhance superconductivity.
  We simulate light-induced phonon dynamics on the anharmonic lattice potential determined by first-principles calculations.
  We find that the anion height $h$, a key structural parameter in iron-based superconductors, approaches its ideal value when an appropriate infrared-active phonon mode is selectively excited.
  This result suggests the possibility of controlling crystal structures and enhancing superconductivity in iron-based superconductors based on the concept of nonlinear phononics.
\end{abstract}
\maketitle

\section{Introduction\label{sec_intro}}

Over the past decade, remarkable progress in generating and controlling coherent phonons has opened new avenues for exploring and engineering nonequilibrium lattice dynamics~\cite{Y_Zhai_2024}.
Although electronic excitation has been widely employed as a trigger for generating coherent phonons, an alternative pathway is phonon-driven ionic Raman scattering~\cite{A_Maradudin_1970,R_Wallis_1971}.
In ionic Raman scattering, resonant optical excitation of an infrared-active (IR) lattice vibration induces a nonlinear displacement of a Raman mode through anharmonic IR-Raman coupling, leading to a transient modification of a crystal structure.
This mechanism is a central concept of nonlinear phononics, a route to controlling crystal structures through light-induced phonon excitation~\cite{M.Forst_2011,M_Forst_2011_2, M_Forst_2013,A.Subedi_2015,R.Mankowsky_2017,M.Fechner_2024,C.Paiva_2024,D.M.Juraschek_2017,G.Mingqiang_2018,A.S.Disa_2020,Y.Zeng_2023,R.Tang_2023,T.G.Blank_2023,A.Subedi_2014,R.Mankowsky_2014,R_Mankowsky_2015,M.Fechner_2016,phono_cavity,phono_review_disa,phono_review_comp,S_Kamiyama_2025}.
Although structural changes in crystals are transient, much longer timescale of phonons than that of electrons allows us to investigate how such structural changes affect electronic states of materials.
Recent progress in intense mid-infrared and terahertz (THz) pulse generation has experimentally enabled nonlinear phononics, and improved probe methods have made it possible to track ultrafast dynamics of various materials~\cite{M.Forst_2011,phono_review_disa}.
Nonlinear phononics has been applied to ferroelectrics~\cite{R.Mankowsky_2017,M.Fechner_2024,A.Subedi_2015,C.Paiva_2024}, magnetic materials~\cite{D.M.Juraschek_2017,A.S.Disa_2020,G.Mingqiang_2018,C.Paiva_2024,Y.Zeng_2023}, topological materials~\cite{R.Tang_2023,T.G.Blank_2023}, and metal-to-insulator transitions~\cite{M_Rini_2007,M.Forst_2011, M_Forst_2011_2, M_Forst_2013, A.Subedi_2014,R_Mankowsky_2015}

In high-$T_c$ superconductors, experimental observations and theoretical studies based on the concept of nonlinear phononics have also been reported.
For example, in the light-irradiated cuprate \ce{YBa2Cu3O_{6.5}} with a mid-infrared pulse, measurements of its optical conductivity revealed superconducting-like behavior at temperatures above $T_c$~\cite{W.Hu_2014,S.Kaiser_2014}, giving rise to discussions on the possibility of photoinduced superconductivity.
Although its driven nonequilibrium state remains under debate~\cite{S_Fava_2024,M_Rosenberg_2025}, one proposed mechanism for a transient structural modification is that ionic Raman scattering induces a displacement of the apical oxygen~\cite{R.Mankowsky_2014,M.Fechner_2016,A.Subedi_2014}.
Meanwhile, we have theoretically proposed the possibility of optical control of the crystal structure of the bilayer nickelate superconductor \ce{La3Ni2O7} based on nonlinear phononics~\cite{S_Kamiyama_2025}.
Since anion positions are crucial for superconductivity in both cuprates~\cite{H_Sakakibara_2010} and bilayer nickelates~\cite{H_Sun_2023}, nonlinear phononics offers a promising route to control superconducting states via anion displacements.

\begin{figure}[t]
  \centering
  \includegraphics[width=0.8\linewidth]{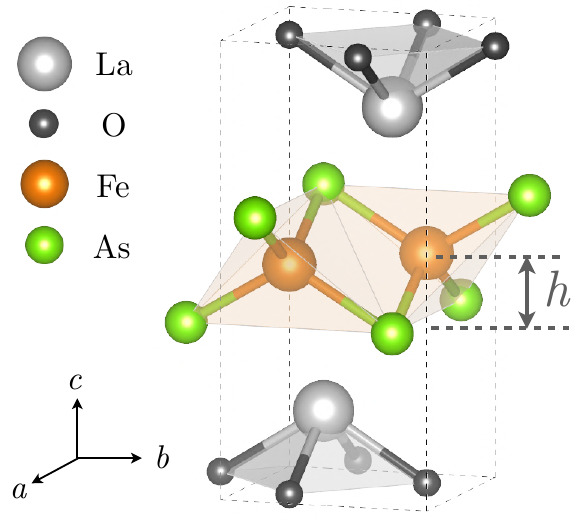}
  \caption{
  Crystal structure of \ce{LaFeAsO} depicted using VESTA~\cite{vesta}. 
  $h$ is the anion height in the Fe--As tetrahedron.
  }
  \label{fig_cryst}
\end{figure}

Iron-based superconductors~\cite{H_Hosono_2015} are interesting targets for nonlinear phononics from the perspective of anion position control.
They comprise various material families, such as the 1111-type, e.g., LaFeAsO~\cite{Y_Kamihara_2008}, the 122-type, e.g., \ce{BaFe2As2}~\cite{M_Rotter_2008}, and the 11-type, e.g., FeSe~\cite{F_Hsu_2008}.
All of these materials have Fe-centered tetrahedra coordinated by anions.
In LaFeAsO, the crystal structure consists of Fe--As layers and La--O layers, as shown in Fig.~\ref{fig_cryst}.
In the superconducting state, the former is responsible for superconductivity, while the latter acts as an insulating layer.

In iron-based superconductors, there is a strong relationship between the local crystal structure and the superconducting transition temperature $T_c$~\cite{C_Lee_2008,K_Kuroki_2009,Y_Mizuguchi_2010}, which is highly sensitive to slight changes in the anion height $h$ of a tetrahedron (see Fig.~\ref{fig_cryst}).
Many experiments have shown that $T_c$ reaches its maximum value around $h=1.38$~\AA~\cite{Y_Mizuguchi_2010}, which corresponds to the anion height in SmFeAsO, one of the iron-based superconductors with the highest $T_c$~\cite{R_ZhiAn_2008}.

Possible photoinduced superconductivity has been suggested in iron-based superconductors through the excitation of coherent phonons mediated by electronic excitation~\cite{K_Okazaki_2018,T_Suzuki_2019,K_Isoyama_2021}.
Displacive excitation of coherent phonons (DECP) is one of the mechanisms in which coherent phonons are generated by photoexcited electrons~\cite{H_Zeiger_1992}.
A transition to an excited electronic state modifies the potential energy surface, leading to a shift of the potential minimum and a corresponding structural distortion.
The slight atomic displacement induced by DECP has been observed in various iron-based superconductors, such as \ce{BaFe2As2}~\cite{S_Gerber_2015,L_Rettig_2015,K_Okazaki_2018} and \ce{FeSe_{1-$x$}Te_{$x$}}~\cite{T_Suzuki_2019,K_Isoyama_2021}.
In particular, in FeSe ($T_{c} \sim 9$~K)~\cite{F_Hsu_2008,S_Kasahara_2014}, superconducting-like behavior was observed under DECP at a temperature of $T=15~\textrm{K}$, above $T_c$, suggesting the possibility of enhanced superconductivity in a nonequilibrium state~\cite{T_Suzuki_2019}.

Based on the above considerations, we focus on the 1111-type iron-based superconductor LaFeAsO for theoretical predictions of crystal-structure control based on nonlinear phononics.
In this study, we aim to bring its structure closer to that of SmFeAsO by inducing ionic Raman scattering with light irradiation, instead of atomic substitution or DECP.
Furthermore, we discuss the possibility of enhancing superconductivity after light irradiation.

This paper is organized as follows.
In Sec.~\ref{sec_opt}, we describe the first-principles calculation methods and present the results of the structural optimization.
In Sec.~\ref{sec_harmonic}, we classify phonon modes based on the crystal symmetry.
In Sec.~\ref{sec_pot}, we construct the anharmonic lattice potential.
In Sec.~\ref{sec_dynamics}, we present the methods and results for calculating phonon dynamics under optical excitation of each IR mode, and identify the most suitable mode for increasing the anion height $h$.
In Sec.~\ref{sec_elec}, we perform first-principles band calculations using the time-averaged crystal structure after optical excitation and investigate the resulting changes in the electronic states.
Discussions are presented in Sec.~\ref{sec_discussions}.
Finally, Sec.~\ref{sec_conclusion} summarizes our study.

\section{structural optimization\label{sec_opt}}
We optimize the crystal structure using first-principles calculations. First-principles calculations based on the density functional theory (DFT) are performed by using the PBE-GGA exchange-correlation functional~\cite{ref_PBE} as implemented in Vienna {\it ab initio} simulation package (VASP)~\cite{vasp1,vasp2,vasp3,vasp,vasp4}.  We use a plane-wave cutoff energy of 600 eV for Kohn-Sham orbitals and a $20\times 20\times 16$ Monkhorst-Pack ${\bm k}$ mesh.

We perform structural optimization until the Hellmann-Feynman force becomes less than 0.01 eV \AA$^{-1}$ for each atom. 
Both the lattice parameters and atomic coordinates are optimized for the tetragonal structure (space group:~$P4/nmm$).
In iron-based superconductors, there is a mismatch between the crystal structures obtained by first-principles calculations and that measured by experiments~\cite{I_Mazin_2008}.
Therefore, we first evaluate $h_{\ce{La}}$, the anion height of LaFeAsO (the target of this study), and $h_{\textrm{Sm}}$, that of SmFeAsO (exhibiting the highest $T_c$).
Then we compare the difference $\Delta h_{\ce{La}-\ce{Sm}} = h_{\ce{Sm}}-h_{\ce{La}}$ obtained from the calculations with that from experiments.
The results are shown in Table~\ref{table_opt}.
Although the values of $h$ differ between the calculated and experimental structures, the relative differences $\Delta h_{\ce{La}-\ce{Sm}}$ shows good agreement.
We then define $h_{\ce{Sm}}^{\textrm{calc.}}=1.227$~\AA , the anion height in the theoretically optimized structure of SmFeAsO, as the target value for the light-induced structural change.

\begin{table}[h]
  \centering
  \begin{tabular}{ccc}
  \hline\hline
    $h$~(\AA)  & calc. & exp. \\ \hline
    $h_{\ce{La}}$ & 1.192 & 1.321(2) \\ 
    $h_{\ce{Sm}}$ & 1.227 & 1.368(3) \\ 
    $\Delta h_{\ce{La}-\ce{Sm}}$ & 0.035 & 0.047(4) \\ \hline\hline
    \end{tabular}
  \caption{
  Anion height $h$ of LaFeAsO ($h_{\ce{La}}$) and SmFeAsO ($h_{\ce{Sm}}$) in both calculated and experimental structures.
  $\Delta h_{\ce{La}-\ce{Sm}}$ is the difference in $h$ between the two materials, i.e., $\Delta h_{\ce{La}-\ce{Sm}} = h_{\ce{Sm}}-h_{\ce{La}}$.
  The experimental values are taken from Refs.~\cite{Y_Kamihara_2008,A_Martinelli_2008}.
  }
  \label{table_opt}
\end{table}

\section{Harmonic phonon calculation\label{sec_harmonic}}
\begin{figure*}[tbp]
  \centering
  \includegraphics[width=1.0\linewidth]{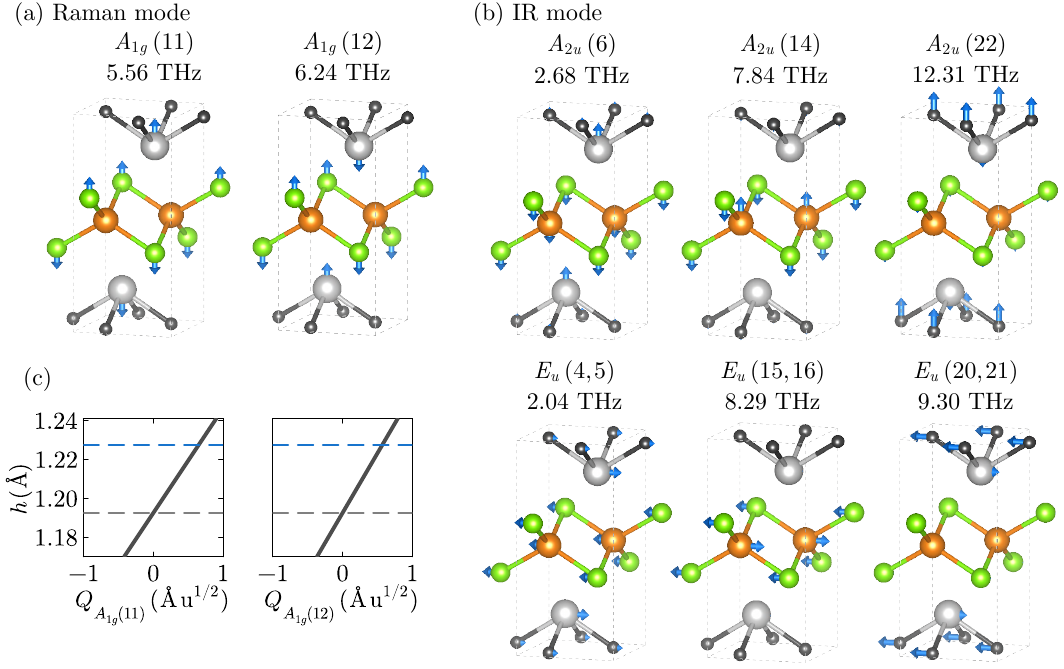}
  \caption{(a)(b) Eigenmodes of LaFeAsO at the $\Gamma$ point (illustrated with VESTA~\cite{vesta}). (a) $A_{1g}$ modes selected as the target Raman modes and (b) $A_{2u}$ and $E_u$ modes selected as the target IR modes. Each label indicates the order counted from the lowest frequency mode.
  (c) Relationship between the phonon normal coordinate $Q$ and the anion height $h$ in the Raman $A_{1g}$ modes.
  The gray and blue dashed lines indicate the theoretical values of $h$ in equilibrium states for LaFeAsO and SmFeAsO, respectively.
}
\label{fig_eigenmodes}
\end{figure*}
Next, we perform harmonic phonon calculations to determine the eigenmodes and eigenfrequencies.
Phonon calculation is performed using the frozen-phonon method as implemented in the Phonopy package~\cite{phonopy_1, phonopy_2}. 
To describe optical responses, we only calculate the $\Gamma$ point phonons~\cite{A.Subedi_2014,R.Mankowsky_2014}. The irreducible representations and frequencies of all eigenmodes are presented in Appendix~\ref{appendix_group}.

Then, we analyze the irreducible representation of each phonon mode for the point group $4/mmm$. 
At the $\Gamma$ point, the $A_{1g}$, $B_{1g}$, and $E_g$ phonon modes are classified as Raman modes, while the $A_{2u}$ and $E_u$ phonon modes are classified as IR modes. 
The former have even parity, whereas the latter have odd parity with respect to space inversion symmetry. 
For Raman modes, since our focus is on controlling the $z$ coordinate of As, we consider only phonon modes that vibrate along the $c$ axis. 
The $A_{1g}$ mode induces a displacement of As atoms along the $c$ axis, which uniformly changes the anion height $h$ in all Fe-As tetrahedra, as seen in Fig.~\ref{fig_eigenmodes}(a).
Owing to its fully symmetric character, the $A_{1g}$ mode preserves the crystal symmetry while changing the position of atoms.
Therefore, we restrict Raman modes considered in this study to the $A_{1g}$ modes.
Note that, as explained in Appendix~\ref{appendix_group}, $B_{1g}$ modes also have nonzero third-order coupling with some IR modes.
Since displacements of As atoms are forbidden for $B_{1g}$ modes by symmetry, we do not consider $B_{1g}$ modes in the following analysis.
The contribution from the $B_{1g}$ modes is discussed in Sec.~\ref{subsec_B1g}.

All IR and $A_{1g}$ Raman modes are shown in Fig.~\ref{fig_eigenmodes}. 
In the IR modes, the one-dimensional representation $A_{2u}$ corresponds to a mode oscillating along the $c$ axis, whereas the two-dimensional $E_u$ modes correspond to oscillations within the $ab$ plane. 
Here, we take two eigenmodes that oscillate along the $a$ and $b$ axes, for doubly degenerate $E_u$ modes.

\section{anharmonic lattice potential\label{sec_pot}}
\subsection{Formulation of lattice potential\label{subsec_pot_method}}
We construct anharmonic lattice potentials based on first-principles calculations.
In this study, we consider anharmonicity up to the fourth order.
The IR and Raman mode potentials, $U_{\textrm{IR}}$ and $U_{\textrm{R}}$, are expressed in terms of their normal coordinates, $Q_{\textrm{IR}}$ and $Q_{\textrm{R}}$, as follows:
\begin{subequations}
  \label{eq:terms}
  \begin{align}
    U_{\textrm{IR}}(Q_{\textrm{IR}}) &= 
    \frac{1}{2}\omega_{\textrm{IR}}^2Q_{\textrm{IR}}^2 
    +a_{4;\textrm{IR}}Q_{\textrm{IR}}^4, \\
    U_{\textrm{R}}(Q_{\textrm{R}}) &= 
    \frac{1}{2}\omega_{\textrm{R}}^2Q_{\textrm{R}}^2 
    +a_{3;\textrm{R}}Q_{\textrm{R}}^3
    +a_{4;\textrm{R}}Q_{\textrm{R}}^4.
  \end{align}
\end{subequations}
$\omega_{\textrm{IR}}$ and $\omega_{\textrm{R}}$ are the phonon frequencies of the IR and Raman modes, respectively, which are obtained by first-principles phonon calculations (shown in Fig.~\ref{fig_eigenmodes}).
$a_{3;\rm{R}}$ and $a_{4;\rm{IR/R}}$ are the third- and fourth-order anharmonic constants of the IR/Raman modes, respectively.
Here, since the IR mode has odd parity, the odd-order anharmonic terms in $U_\textrm{IR}$ are forbidden by the space inversion symmetry.

$a_{3;\rm{R}}$ and $a_{4;\rm{R}}$ are determined by the fourth-order polynomial fitting using the DFT total energies for various $Q_{\rm{R}} \in [-2.0,2.0]$~\AA~$\rm{u}^{1/2}$ with $Q_{\rm{IR}}=0$.
Here, u denotes the atomic mass unit.
The same fitting is performed for $a_{4;\rm{IR}}$ using the DFT total energies for various $Q_{\rm{IR}}\in [-2.0,2.0]$~\AA~$\rm{u}^{1/2}$ with $Q_{\rm{R}}=0$.

\subsubsection{$A_{2u}-A_{1g}$ coupling \label{subsec_pot_method_A2u}}
Let us consider the anharmonic lattice potential for one IR and two Raman $A_{1g}$ modes.
Hereafter, we refer to the two $A_{1g}$ Raman modes $A_{1g}(11)$ and $ A_{1g}(12)$ as $\textrm{R}_1$ and $\textrm{R}_2$, for simplicity.

First, we consider the case where an $A_{2u}$ phonon mode is taken as the IR mode for photoexcitation.
Including anharmonic terms up to the fourth order that preserve the space inversion symmetry, the lattice potential is expressed as
\begin{align}
  &V_{A_{2u}-A_{1g}}(Q_{\textrm{IR}},Q_{\textrm{R}_1},Q_{\textrm{R}_2})\notag\\
  &=U_{\textrm{IR}}(Q_{\textrm{IR}}) + \sum_{i}U_{\textrm{R}}(Q_{\textrm{R}_i}) \label{eq:pot_A2u}  \\
  &-\frac{1}{2}\sum_{i} g_{\textrm{IR}-\textrm{R}_i} Q_{\textrm{IR}}^2Q_{\textrm{R}_i}
  -\frac{1}{2}\sum_{i}h_{\textrm{IR}-\textrm{R}_i} Q_{\textrm{IR}}^2Q_{\textrm{R}_i}^2, \notag
\end{align}
where $i=1,2$.
$g_{\rm{IR}-\rm{R}}$ and $h_{\rm{IR}-\rm{R}}$ are the third- and fourth-order IR-Raman coupling constants, respectively.
Here, we neglect Raman-Raman coupling terms~(see discussion in Sec.~\ref{subsec_lifetime}).

Once $U_{\textrm{IR}}(Q_{\textrm{IR}})$ and $U_{\textrm{R}}(Q_{\textrm{R}})$ are known, the parameters $g_{\rm{IR}-\rm{R}}$ and $h_{\rm{IR}-\rm{R}}$ can be determined as follows. 
First, the DFT total energies are computed for $Q_{\rm{IR}}=1.0$ and various $Q_{\rm{R}_1}\in [-2.0,2.0]$ \AA~$\rm{u}^{1/2}$, with fixing $Q_{\textrm{R}_2}=0$.
Then, $U_{\textrm{IR}}(Q_{\textrm{IR}})$ and $U_{\textrm{R}}(Q_{\textrm{R}_1})$ are subtracted from $V_{A_{2u}-A_{1g}}$, and the remaining energy is fitted with a second-order polynomial in $Q_{\rm{R}_1}$ to extract $g_{\rm{IR}-\rm{R}_1}$ and $h_{\rm{IR}-\rm{R}_1}$.
Same procedure is performed to obtain $g_{\rm{IR}-\rm{R}_2}$ and $h_{\rm{IR}-\rm{R}_2}$.

\subsubsection{$E_{u}-A_{1g}$ coupling}
Next, we consider the case where $E_u$ modes are chosen as the IR modes.
As explained in Sec.~\ref{sec_harmonic}, since $E_u$ is a two-dimensional irreducible representation, its eigenmodes that oscillate along orthogonal directions are degenerate.
We denote the normal coordinates of phonon modes that oscillate along the $a$ and $b$ axes as $Q_{\textrm{IR},a}$ and $Q_{\textrm{IR},b}$, respectively.
The lattice potential is given by
\begin{align}
  &V_{E_{u}-A_{1g}}(Q_{\textrm{IR},a},Q_{\textrm{IR},b},Q_{\textrm{R}_1},Q_{\textrm{R}_2})\notag \\
  &=\sum_{\alpha} U_{\textrm{IR}}(Q_{\textrm{IR},\alpha}) + \sum_{i}U_{\textrm{R}}(Q_{\textrm{R}_i}) \notag \\
  &-\frac{1}{2}\sum_{i}g_{\textrm{IR}-\textrm{R}_i} (Q_{\textrm{IR},a}^2+Q_{\textrm{IR},b}^2)Q_{\textrm{R}_i}\label{eq:pot_Eu} \\
  &-\frac{1}{2}\sum_{i}h_{\textrm{IR}-\textrm{R}_i} (Q_{\textrm{IR},a}^2+Q_{\textrm{IR},b}^2)Q_{\textrm{R}_i}^2 \notag \\
  &-\frac{1}{2} \tilde{h}_{\textrm{IR}-\textrm{IR}}Q_{\textrm{IR},a}^2Q_{\textrm{IR},b}^2 ,\notag
\end{align}
where $\alpha \in \{a, b\}$, and $\tilde{h}_{\textrm{IR}-\textrm{IR}}$ is the IR-IR coupling constant.

Regarding the IR-Raman coupling terms, since the Raman mode belongs to the $A_{1g}$ irreducible representation and the point group $4/mmm$ has $C_4$ symmetry, the coefficients of the third-order coupling terms $Q_{\textrm{IR},a}^2 Q_{\textrm{R}}$ and $Q_{\textrm{IR},b}^2 Q_{\textrm{R}}$ are identical, and the coefficient of $Q_{\textrm{IR},a} Q_{\textrm{IR},b} Q_{\textrm{R}}$ vanishes. 
Therefore, the third-order coupling terms have the form $(Q_{\textrm{IR},a}^2+Q_{\textrm{IR},b}^2)Q_{\textrm{R}}$.
Similarly, the fourth-order coupling terms have the form $(Q_{\textrm{IR},a}^2+Q_{\textrm{IR},b}^2)Q_{\textrm{R}}^2$.
The IR-IR coupling term of the form $Q_{\textrm{IR},a}^2Q_{\textrm{IR},b}^2$ is allowed due to the space inversion symmetry and the mirror symmetries about the $ac$ and $bc$ planes.

We determine the coefficients of Eq.~(\ref{eq:pot_Eu}) as follows.
We set $Q_{\textrm{IR},b}=0$ and determine $g_{\textrm{IR}-\textrm{R}_i}$ and $h_{\textrm{IR}-\textrm{R}_i}$ for a pair of $(Q_{\textrm{IR},a},Q_{\textrm{R}_i})$ $(i=1,2)$ using the same procedure as that described in Sec.~\ref{subsec_pot_method_A2u}.
We then set $Q_{\textrm{R}_1}=Q_{\textrm{R}_2}=0$ and determine $\tilde{h}_{\textrm{IR}-\textrm{IR}}$ over the ranges $Q_{\textrm{IR},a}\in [-2.0,2.0]$~\AA~$\rm{u}^{1/2}$ and $Q_{\textrm{IR},b}\in [-2.0,2.0]$~\AA~$\rm{u}^{1/2}$.

\subsection{Results: Fitted potential and coupling constants}
Based on the procedure described in Sec.~\ref{subsec_pot_method}, we determine the coefficients of anharmonic lattice potentials, Eqs.~(\ref{eq:pot_A2u}) and (\ref{eq:pot_Eu}), for all pairs of IR and Raman modes considered in this study.

\begin{figure}[h]
  \centering
  \includegraphics[width=0.9\linewidth]{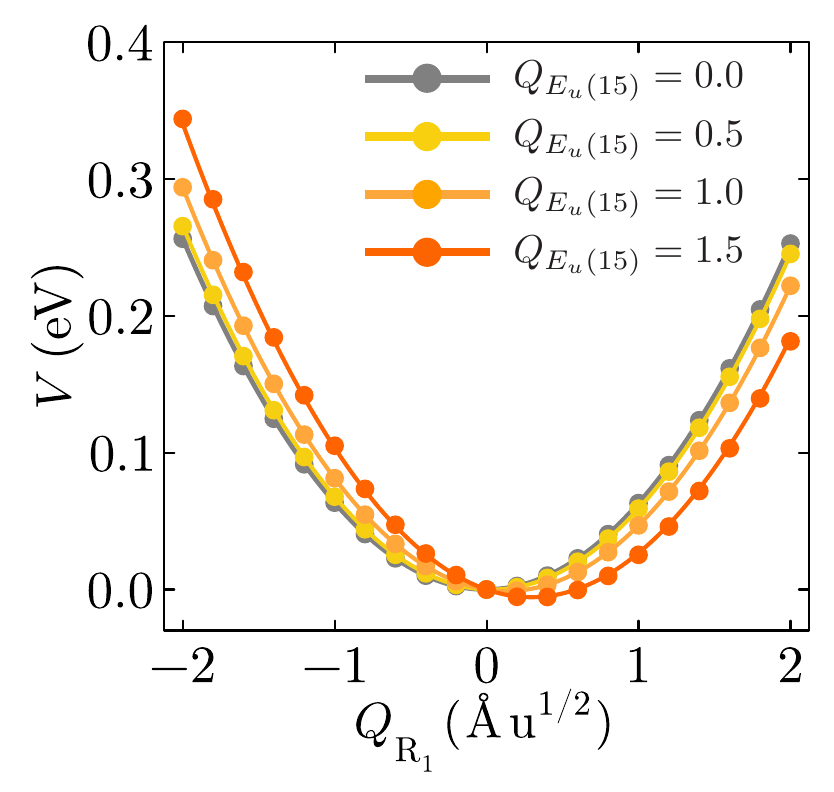}
  \caption{Anharmonic lattice potential $V_{E_u-A_{1g}}(Q_{E_u(15)},\allowbreak Q_{E_u(16)}=0,\allowbreak Q_{\textrm{R}_1},\allowbreak Q_{\textrm{R}_2}=0)$ for the IR mode $E_u(15)$ and the Raman mode $\textrm{R}_1$, where circles represent DFT total energies and curves represent lattice potentials fitted by Eq.~(\ref{eq:pot_Eu}).}
  \label{fig_pot_curve}
\end{figure}

As an example, Fig.~\ref{fig_pot_curve} shows the DFT total energies for various displacements of the $E_u(15)$ and $\textrm{R}_1$ modes with the fitted curve $V_{E_u-A_{1g}}(Q_{E_u(15)},Q_{E_u(16)}=0,Q_{\textrm{R}_1},Q_{\textrm{R}_2}=0)$~\footnote{
The $E_u(15)$ and $E_u(16)$ modes correspond to displacements along the $a$ and $b$ axes, respectively.
}.
The result demonstrates that the fitted potential accurately reproduces the DFT total energies.
For $Q_{E_u(15)}\neq 0$, the minimum of the potential curve is located at $Q_{\textrm{R}_1}\neq 0$, indicating a shift of the equilibrium position.

\begin{figure}[h]
  \centering
  \includegraphics[width=1.0\linewidth]{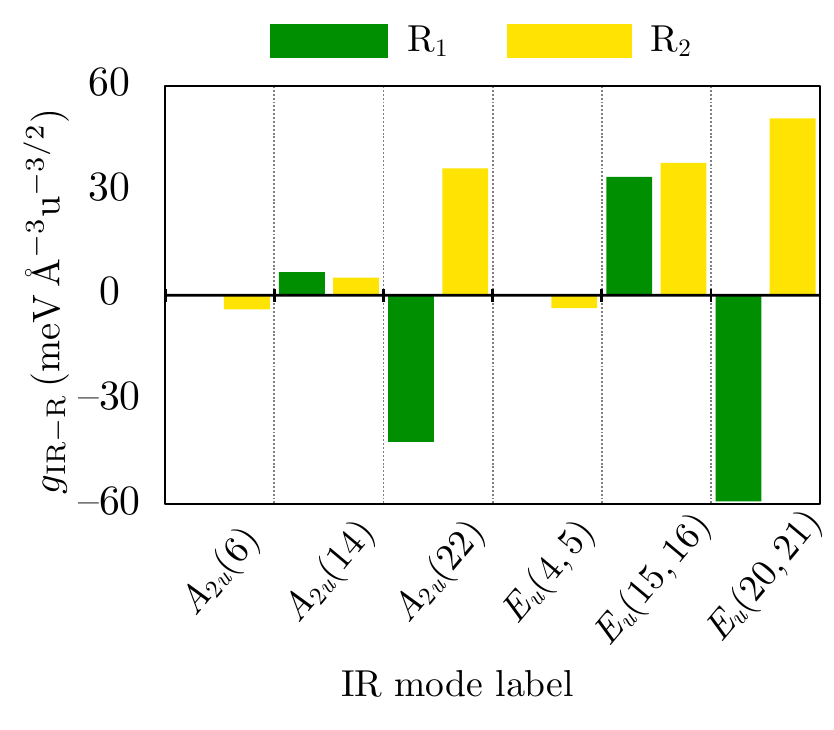}
  \caption{Values of the third anharmonic IR-Raman coupling $g_{\textrm{IR}-\textrm{R}}$. The horizontal axis corresponds to the IR mode label.
  The green and yellow bars represent the Raman $\textrm{R}_1$ and $\textrm{R}_2$ modes, respectively.}
  \label{fig_coupling_IR_Raman}
\end{figure}

Figure~\ref{fig_coupling_IR_Raman} presents the third-order coupling constants $g_{\text{IR}-\text{R}}$ for each pair of IR and Raman modes.
Here, if the sign of $g_{\textrm{IR}-\textrm{R}}$ is positive, the potential minimum shifts to $Q_{\textrm{R}}>0$ when $Q_{\textrm{IR}}\neq 0$.
Similarly, if $g_{\textrm{IR}-\textrm{R}}$ is negative, the potential minimum shifts to $Q_{\textrm{R}}<0$.
Therefore, the sign of $g_{\textrm{IR}-\textrm{R}}$ coincides with the sign of $Q_{\textrm{R}}$ at the equilibrium position.
A change in the anion height $\Delta h$ becomes positive ($\Delta h>0$) when the Raman modes has $Q_{\textrm{R}}>0$, as shown in Fig.~\ref{fig_eigenmodes}(c), which illustrates the relationship between $Q_{\textrm{R}}$ and $h$.
As the absolute value of $g_{\textrm{IR}-\textrm{R}}$ is large, the amplitude of the induced Raman mode becomes large.
Consequently, a large positive $g_{\textrm{IR}-\textrm{R}}$ may lead to a large positive $\Delta h$.

In the case of the IR $E_u(15,16)$ mode, the couplings to both Raman ($A_{1g}$) modes are positive ($g_{\textrm{IR}-\textrm{R}}>0$) and have large absolute values.
Therefore, resonant excitation of the $E_u(15,16)$ mode can lead to an increase in the anion height $h$ owing to the large positive IR-Raman couplings $g_{\textrm{IR}-\textrm{R}}$.
For the $A_{2u}(14)$ mode, the values of $g_{\textrm{IR}-\textrm{R}}$ are positive for both Raman modes, although their absolute values are small, which may still lead to a slight increase in $h$.
On the other hand, the $A_{2u}(22)$ and $E_{u}(20,21)$ modes have large $|g_{\textrm{IR}-\textrm{R}}|$ even though one of the two coupling constants, $g_{\textrm{IR}-\textrm{R}_1}$, is negative.

The $A_{2u}(22)$ and $E_{u}(20,21)$ modes are eigenmodes associated with atomic oscillations in the LaO layer.
This likely leads to large $g_{\textrm{IR}-\textrm{R}}$ values through coupling to the La atomic components of the eigenvectors of $\textrm{R}_1$ and $\textrm{R}_2$~\footnote{
  The different sign of $g_{\textrm{IR}-\textrm{R}}$ between $\textrm{R}_1$ and $\textrm{R}_2$ modes likely originates from the difference in the directions of La atomic displacements between the two Raman modes.
}.
In this material, the $E_u$ and $A_{1g}$ modes correspond to oscillations in the $ab$ plane and along the $c$ axis, respectively.
Although the displacement vectors of the $E_u$ and $A_{1g}$ modes are perpendicular to each other at each atom, the IR-Raman coupling constants take large values.
In contrast, for the $A_{2u}(6)$ and $E_u(4,5)$ modes, the coupling constants $g_{\textrm{IR}-\textrm{R}}$ is much smaller than those of the other modes~\footnote{
  The values of $g_{A_{2u}(6)-\textrm{R}_1}$, $g_{A_{2u}(6)-\textrm{R}_2}$, $g_{E_{u}(4,5)-\textrm{R}_1}$, $g_{E_{u}(4,5)-\textrm{R}_2}$ are $2.48\times 10^{-1}$, $-4.13$, $-2.40\times 10^{-2}$, $-3.71$ meV \AA$^{-3}$ u$^{-3/2}$, respectively.}.
Since the $A_{2u}(6)$ and $E_u(4,5)$ modes are not expected to induce an increase in $h$, we exclude them from the subsequent analysis.

\section{Nonlinear phonon dynamics and modification of the crystal structure\label{sec_dynamics}}
\subsection{Equation of motion}
In this section, we calculate the time evolution of the phonon modes and dynamics of the crystal structure after optical excitation.
In this study, we consider the case when a specific IR mode is resonantly excited by the optical pulse field, and Raman oscillations are induced through the IR-Raman coupling.
Here, we assume that both of the two $A_{1g}$ modes are simultaneously activated by a single IR mode.
Note that we treat the phonon dynamics classically, as in previous studies~\cite{A.Subedi_2014,S_Kamiyama_2025}.

\subsubsection{$A_{2u}-A_{1g}$ coupling}
First, we consider the case where the $A_{2u}$ mode is optically excited.
The equations of motions for one IR $A_{2u}$ mode and two Raman $A_{1g}$ modes are given by
\begin{subequations}
  \label{eq:eqm_A2u}
  \begin{align}
    \ddot{Q}_{\textrm{IR}} &= F(t) -\frac{\partial V_{A_{2u}-A_{1g}}}{\partial Q_{\textrm{IR}}}\\
    \ddot{Q}_{\textrm{R}_i}&= -\frac{\partial V_{A_{2u}-A_{1g}}}{\partial Q_{\textrm{R}_i}} \quad (i=1,2),
  \end{align}
\end{subequations}
where $F(t) = F_0 e^{-t^2/2\sigma^2}\cos\Omega\,t$ is the driving force generated by the external light pulse. 
We assume that the field is polarized along the $c$ axis, i.e., along the eigenvector direction of the $A_{2u}$ mode~\footnote{
  If the external field is applied in a direction not parallel to the $c$ axis, the projection of its amplitude onto the $c$ axis plays the role of $F_0$ in Eq.~(\ref{eq:eqm_A2u}).
}.
The FWHM (full width at half maximum) of the Gaussian function is represented as $\tau = 2\sqrt{2\ln 2}\sigma$. We set $F_0 = 60\,\rm{meV} \mbox{\AA}^{-1} \rm{u}^{-1/2}$ and $\tau = 0.3$ ps unless otherwise noted. The field frequency $\Omega$ is set as $\Omega = \omega_{\rm{IR}}$, where the IR phonon is resonantly excited.
We set the initial values as $Q=0$ and $\dot{Q}=0$ for all modes.
Note that damping factors are not considered in the equations of motions in the main text (see Appendix~\ref{appendix_damping}). 

\subsubsection{$E_{u}-A_{1g}$ coupling}
Second, we consider the case where the $E_u$ mode is optically excited.
The equations of motions for the normal coordinates of the $E_u$ mode oscillating along the $a$ and $b$ axes ($Q_{\textrm{IR},a}$ and $Q_{\textrm{IR},b}$) and for the two $A_{1g}$ Raman modes are given by
\begin{subequations}
  \label{eq:eqm_Eu} 
  \begin{align}
    \ddot{Q}_{\textrm{IR},a} &= F(t)\cos\theta -\frac{\partial V_{E_{u}-A_{1g}}}{\partial Q_{\textrm{IR},a}}\\
    \ddot{Q}_{\textrm{IR},b} &= F(t)\sin\theta -\frac{\partial V_{E_{u}-A_{1g}}}{\partial Q_{\textrm{IR},b}}\\
    \ddot{Q}_{\textrm{R}_i}&= -\frac{\partial V_{A_{2u}-A_{1g}}}{\partial Q_{\textrm{R}_i}}\quad (i=1,2).
  \end{align}  
\end{subequations}
Here, we assume that the direction of the field lies in the $ab$ plane, and $\theta$ denotes the irradiation angle of $F(t)$.
When $\theta = 0$, the field is polarized along the $a$ axis, and thus $Q_{\textrm{IR},b}(t)=0$; in this case, the equation of motion reduces to Eq.~(\ref{eq:eqm_A2u}).
Here, we set $\theta=0$ in the calculation of phonon dynamics, since the $E_u-A_{1g}$ coupling shows little dependence on the irradiation angle $\theta$, as shown in Appendix~\ref{appendix_theta}.

\subsubsection{Atomic Displacements}
The relationship between the $\Gamma$ point phonon normal coordinate $Q_{\alpha}$ for mode $\alpha$ and the displacement vector $\bm{U}_{\alpha,j}$ for atom $j$ is expressed as
\begin{align}
\label{angle} 
\bm{U}_{\alpha,j} = \frac{Q_{\alpha}}{\sqrt{m_j}} \bm{e}_{\alpha,j},
\end{align} 
where $m_j$ is the mass of atom $j$ and $\bm{e}_{\alpha,j}$ is the dimensionless eigenvector of the phonon mode.

\subsection{Results: Time evolution of phonon modes}
\begin{figure*}[p]
  \centering
  \includegraphics[width=0.95\linewidth]{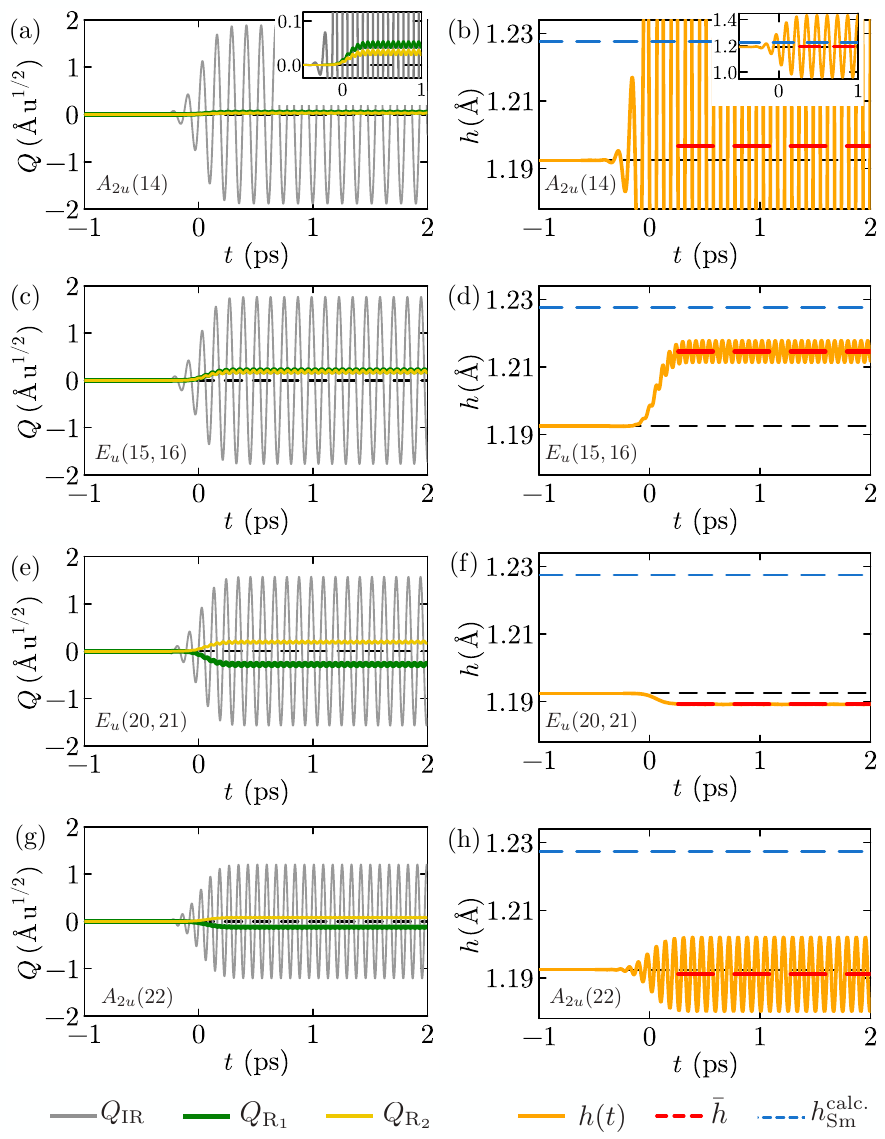}
  \caption{
    Time evolution of the phonon normal coordinate $Q$ (left panels) and the anion height $h$ (right panels).
    Each pair of panels corresponds to resonant excitation of the IR modes: (a)(b)$A_{2u}(14)$, (c)(d)$E_u(15,16)$, (e)(f)$E_u(20,21)$, and (g)(h)$A_{2u}(22)$.
    In the right panels, the red dashed lines represent the time-averaged anion height $\bar{h}$ after light irradiation, and the blue dashed lines show the value of $h$ in the calculated structure of SmFeAsO.
  }
  \label{fig_dynamics_all}
\end{figure*}
Here, we show the time evolution of the phonon modes and the crystal structure under optical excitation of each IR mode.
The obtained time-evolution data of $Q(t)$ and $h(t)$ are averaged over the time windows after the external field is applied, and the averaged values are denoted by $\bar{Q}$ and $\bar{h}$, respectively~\footnote{
  For a time-evolution data $X(t)$, we evaluate a time average of after light irradiation in the time domain $[t_{\mathrm{min}}, t_{\mathrm{max}}]$ as follows:
  \begin{equation*}
  \bar{X} = \frac{1}{t_{\mathrm{max}}-t_{\mathrm{min}}}\int_{t_{\mathrm{min}}}^{t_{\mathrm{max}}} dt' \int_{t_{\mathrm{min}}}^{t'}dt\ \frac{X(t)}{t'-t_{\mathrm{min}}},
  \end{equation*}
  where the time average over $t\in[t_{\mathrm{min}}, t']$ is again time-averaged over $t'\in [t_{\mathrm{min}}, t_{\mathrm{max}}]$.
  While the first time average, $\int_{t_{\mathrm{min}}}^{t'}dt\ \frac{X(t)}{t'-t_{\mathrm{min}}}$, exhibits a decaying oscillation against $t'$, a center of that oscillation is efficiently obtained by the second average.
  We set $t_{\mathrm{min}}=0.8\,\rm{ps}$, which is sufficiently larger than $\tau=0.3\,\rm{ps}$ so that $F(t_{\mathrm{min}})\simeq 0$, and $t_{\mathrm{max}}=25\,\rm{ps}$, which is sufficiently large to get a converged value for the time average.}. 
As mentioned in Sec.~\ref{sec_pot}, the $A_{2u}(6)$ and $E_u(4,5)$ modes are excluded due to their tiny values of $|g_{\textrm{IR}-\textrm{R}}|$.
Figure~\ref{fig_dynamics_all} shows the time evolution of the phonon normal coordinates $Q$ and the anion height $h$ when one IR mode is resonantly excited by light irradiation.

First, we consider the case in which the $A_{2u}(14)$ mode is optically excited [Figs.~\ref{fig_dynamics_all}(a) and \ref{fig_dynamics_all}(b)].
After light irradiation, the oscillation centers of $Q_{\textrm{R}_1}$ and $Q_{\textrm{R}_2}$ slightly shift toward $Q_{\textrm{R}} > 0$.
However, since this IR mode involves oscillations along the $c$ axis, its excitation induces large-amplitude oscillations in $h(t)$.
As a result, although the time-averaged value $\bar{h}$ slightly increases from the equilibrium value, $h(t)$ temporarily decreases below its initial value due to the strong oscillatory component of $Q_{\textrm{IR}}$.

Next, let us turn to the case of optical excitation of the $E_u (15,16)$ mode, which involves atomic oscillations within the $ab$ plane~[Figs.~\ref{fig_dynamics_all}(c) and \ref{fig_dynamics_all}(d)]. 
In this case, the vibrational motion of the IR modes does not directly change $h(t)$.
As a result, the vibrational component in $h(t)$ from the IR mode is suppressed, while $\bar{h}$ exhibits a substantial increase.

A shift of the time-averaged value of $h$ induced by photoexcitation is also observed for the $E_u(20,21)$ case, where the oscillatory component remains negligibly small~[Figs.~\ref{fig_dynamics_all}(e) and \ref{fig_dynamics_all}(f)].
Since the $E_u(20,21)$ mode corresponds to in-plane vibrations of the LaO layer, which are not directly related to the anion height $h$ (see Fig.~\ref{fig_eigenmodes}), the time evolution of $h$ exhibits almost no oscillatory component, and only a shift in its time-averaged value is observed.
It should be noted that although time dependence of $h$ appears not to reflect IR-mode oscillation, vibrational motion occurs within the $ab$ plane.
Similarly, the $A_{2u}(22)$ mode mainly involves oscillations of the LaO layer along the $c$ axis.
As shown in Figs.~\ref{fig_dynamics_all}(g) and \ref{fig_dynamics_all}(h), the oscillation amplitude of $h(t)$ induced by the $A_{2u}(22)$ mode is smaller than that induced by the $A_{2u}(14)$ mode [Fig.~\ref{fig_dynamics_all}(b)].

Since $\bar{h}$ increases the most when stimulating the $E_u(15,16)$ mode, we focus on the case where the $E_u(15,16)$ is chosen as an IR mode hereafter.

\begin{figure}[htbp]
  \centering
  \includegraphics[width=1.0\linewidth]{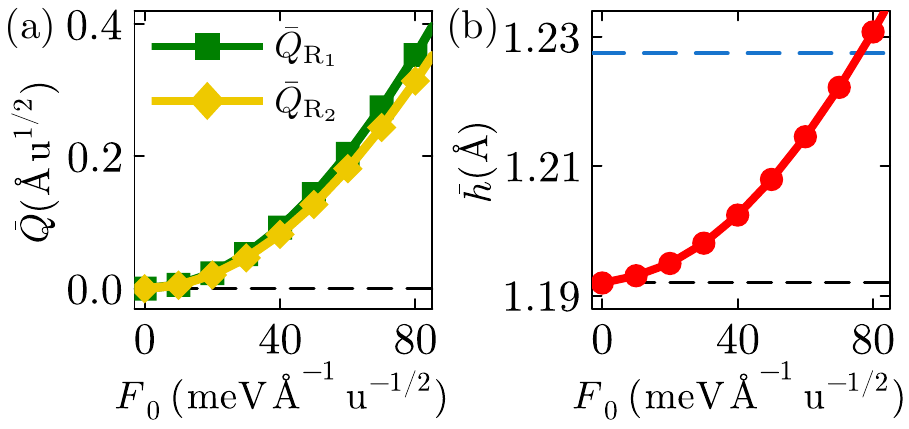}
  \caption{Field-amplitude ($F_0$) dependence of time-averaged values of (a) the Raman-mode displacements $\bar{Q}_{\textrm{R}}$ and (b) the anion height $\bar{h}$ for the case where the $E_u(15,16)$ mode is optically excited.}
  \label{fig_F0}
\end{figure}

The above calculations are performed with the external field amplitude set to $F_0 = 60\,\rm{meV} \mbox{\AA}^{-1} \rm{u}^{-1/2}$.
In Fig.~\ref{fig_F0}, we present how the field amplitude $F_0$ affects the phonon normal coordinates $Q_{\textrm{R}}$ and the anion height $h$.
We find that the changes of $\bar{Q}_{\textrm{R}}$ and $\bar{h}$ are roughly proportional to $F_0^2$, as seen in our previous calculation~\cite{S_Kamiyama_2025}.
This originates from approximate relationships, $Q_{\rm{IR}}\propto F_0$ and $Q_{\rm{R}}\propto Q_{\rm{IR}}^2$, which are because the IR and Raman modes are mainly driven by the external field and the phonon-phonon coupling proportional to $Q_{\rm{IR}}^2$, respectively [see, Eqs.~(\ref{eq:eqm_A2u}) and (\ref{eq:eqm_Eu})].
We here demonstrate the quadratic $F_0$ dependence of the structural changes as a consequence of the nonlinear Raman oscillation, which is a key aspect of nonlinear phononics for realizing a nonzero displacement of the time-averaged crystal structure after light irradiation.
We also note that a change of anion height $h$ can be greatly increased by using high-intensity light.

Note that in iron-based superconductors, one can regard that the Fe-As-Fe bond angle $\alpha$ affects $T_c$, similarly to $h$~\cite{C_Lee_2008}.
Since these two parameters are closely correlated, we consider only the anion height $h$.
$\alpha$ in the time-averaged crystal structure is shown in Appendix~\ref{appendix_alpha}.
Similarly to the anion height $h$, $\alpha$ approaches the value of SmFeAsO obtained in the calculated structure.

\section{Electronic band structure for the time-averaged crystal structure \label{sec_elec}}
\begin{figure}[t]
  \centering
  \includegraphics[width=1.0\linewidth]{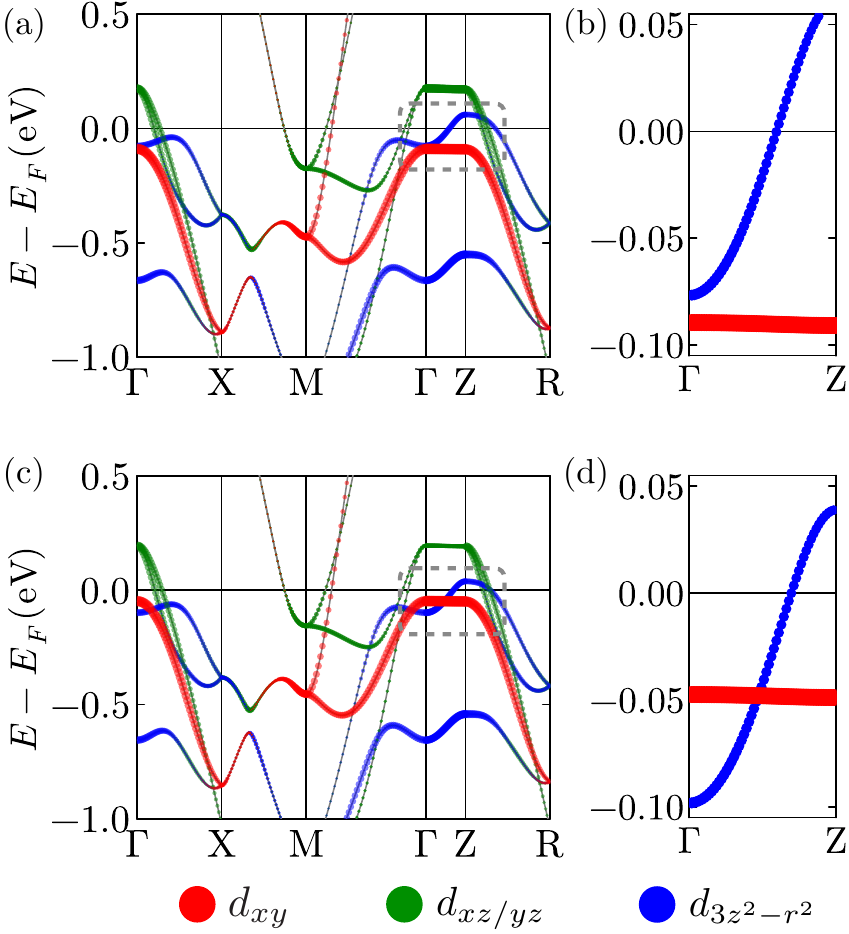}
  \caption{
  Electronic band dispersions for (a)(b) the equilibrium crystal structure before light irradiation and (c)(d) the time-averaged crystal structure after light irradiation.
  The color indicate the orbital characters: red for $d_{xy}$, blue for $d_{3z^2-r^2}$, and green for $d_{xz/yz}$.
  The $x$ and $y$ axes for the $d$ orbitals are aligned with the directions of the nearest-neighbor Fe site, while the $z$ axis corresponds to the $c$ axis.
}
  \label{fig_band}
\end{figure}

In this section, to investigate how the electronic structure is affected by the structural distortion by light irradiation, we calculate the electronic band dispersion for the time-averaged crystal structure after light irradiation, when the $E_u(15,16)$ mode is optically excited. 
In Fig.~\ref{fig_band}, we present the electronic band dispersions for the original (unmodified) crystal structure and those for the modified crystal structure.
Note that the $x$ and $y$ axes for the Fe $3d$ orbitals are defined along the nearest-neighbor Fe-Fe directions (corresponding to a $45^{\circ}$ rotation of the $a$ and $b$ axes), while the $z$ axis is parallel to the $c$ axis.

The most notable effect of the light-induced structural change appears in the band dispersions of the $d_{xy}$ and $d_{3z^2-r^2}$ orbitals around the Fermi level along the $\Gamma-Z$ line.
In the enlarged views~[Figs.~\ref{fig_band}(b) and \ref{fig_band}(d)], the band of the $d_{3z^2-r^2}$ orbital (blue) shifts downward, while the band of the $d_{xy}$ orbital (red) moves closer to the Fermi level.
As explained in Sec.~\ref{sec_dynamics}, the anion height $h$ increases upon optical excitation of the $E_u(15,16)$ mode.
It is well known that increasing $h$ narrows the bandwidth of the $d_{xy}$ orbital and moves close to the Fermi level~\cite{K_Kuroki_2009}.
This is consistent with the change obtained in calculations induced by optical excitation of the $E_u(15,16)$ mode.

The upward shift of the $d_{xy}$ band toward the Fermi level due to an increase in $h$ corresponds to the formation of an additional Fermi surface, which in turn enhances $T_c$ according to a theoretical study~\cite{K_Kuroki_2009}.
It has also been pointed out that superconductivity in iron-based compounds is optimized when the $d_{xy}$ band comes closest to the Fermi level, i.e., in the incipient-band regime~\cite{K_Kuroki_2009,P_Hirschfeld_2011,Y_Bang_2016,M_Nakata_2017,Y_Bang_2019}, which is realized around $h=1.38$~\AA~in SmFeAsO, the material exhibiting the highest $T_c$ among iron-based superconductors~\cite{C_Lee_2008,Y_Mizuguchi_2010}.
Therefore, we expect that the change in the energy bands induced by excitation of the $E_u(15,16)$ can contribute to an enhancement of superconductivity.

Note that all the above calculations are based on structures derived from first-principles structural optimization to get a phonon dispersion with no imaginary modes.
A difference in the band dispersions calculated using the theoretical and experimental crystal structures of LaFeAsO lies in the position of the $d_{3z^2-r^2}$ band~\cite{I_Mazin_2008}.
In the calculated equilibrium crystal structure~[Figs.~\ref{fig_band}(a) and \ref{fig_band}(b)], the $d_{3z^2-r^2}$ band appears to intersect the Fermi level along the $\Gamma-Z$ line, forming a Fermi surface.
However, in the band calculation of the experimental crystal structure, the $d_{3z^2-r^2}$ band does not intersect the Fermi level and therefore does not form a Fermi surface.
Thus, we consider that the $d_{3z^2-r^2}$ band might not intersect the Fermi level even in the experimentally photo-excited electronic state.

We analyze the electronic state after light irradiation by performing band calculations assuming a time-averaged crystal structure.
However, we do not consider the effect of lattice vibrations on the electronic state during the time-evolution, which remains an important issue for future investigation.

\section{Discussions \label{sec_discussions}}
\subsection{Phonon-phonon coupling and lifetime \label{subsec_lifetime}}
In this paper, we include phonon-phonon couplings only between specific modes.
In particular, we consider only IR-Raman coupling at $\bm{q}=\bm{0}$, and IR-IR coupling between degenerated modes.
We neglect the following couplings: 
(i) Raman-Raman coupling at $\bm{q}=\bm{0}$, 
(ii) couplings to $\bm{q}\neq\bm{0}$ modes.

(i) The effect of Raman-Raman couplings to nonlinear phononics is discussed in Ref.~\cite{Raman-Raman}.
The terms like $h_{\rm{R}_1 - \rm{R}_2} Q_{\rm{R}_1}Q_{\rm{R}_2}$ is to renormalize the light-induce phonon displacements. We expect that our results are qualitatively kept unchanged even by including the Raman-Raman coupling, while the size of the Raman displacements can be somewhat renormalized.

(ii) For the effect of couplings to $\bm{q}\neq\bm{0}$ modes:
To include such coupling terms, we should consider intractably many phonon modes over the whole Brillouin zone.
On the other hand, the nonlinear oscillation, which is an objective of our study, requires $Q_{\rm{IR}}^2$ dependence of the potential term.
As mentioned in Sec.~\ref{sec_harmonic}, since IR modes are optically excited, they must have zero crystal momentum ($\bm{q}_{\textrm{IR}}=\bm{0}$).
The lowest-order term that can induce the nonlinear oscillation is $Q_{\rm{IR}}^2Q_{\rm{R}}$, for which only phonons at the $\Gamma$ point are involved due to the translational invariance of the crystal (i.e., the total crystal momentum should be zero for each term).
Thus, in the present study, we ignore ${\bm q}\neq{\bm 0}$ phonons for simplicity as many theoretical studies on nonlinear phononics assumes~\cite{M.Fechner_2016,R.Mankowsky_2014,A.Subedi_2014,R.Tang_2023,T.G.Blank_2023,A.S.Disa_2020,phono_cavity,Y.Zeng_2023,A.Subedi_2015,G.Mingqiang_2018}.

In this work, we neglect damping effects arising from phonon scattering processes that lead to a finite phonon lifetime.
The phonon lifetime arises from electron-phonon coupling and phonon-phonon coupling.
We treat phonon-phonon coupling only as an origin of inducing crystal-structure modifications, neglecting the effect of finite phonon lifetimes.
In fact, when one seeks to see the transient modulated crystal structure after light irradiation theoretically, damping effects are not necessarily included. 
Nonetheless, in some studies, the lifetime was taken as the damping term in the equation of motion phenomenologically~\cite{R.Mankowsky_2014,Y.Zeng_2023}.
We show the phonon dynamics including a damping term in Appendix~\ref{appendix_damping}. 
We confirm that a change in the crystal structure is similar to our simulation presented in Sec.~\ref{sec_dynamics} within a short timescale following the field irradiation.
In any case, it is expected the IR-Raman third-order coupling term $g_{\rm{IR}-\rm{R}}Q_{\rm{IR}}^2Q_{\textrm{R}}$ at $\bm{q}=\bm{0}$ could have the largest effect on structural modification.
Although the other coupling effects may influence the phonon lifetime, this remains one of the future tasks.

\subsection{Effects of $B_{1g}$ modes \label{subsec_B1g}}
In this paper, we consider only the $A_{1g}$ modes as Raman modes.
However, when the $E_u$ mode is optically excited, oscillations of the $B_{1g}$ modes are also induced through anharmonic coupling, as shown in Appendix~\ref{appendix_group}.
Nevertheless, the effect of the $B_{1g}$ modes can be neglected by adjusting the irradiation angle $\theta$, as discussed below.

The lattice potential for a pair of $E_u$ and $B_{1g}$ modes is expressed as
\begin{align}
  \label{eq:pot_B1g}
  &V_{E_{u}-B_{1g}}(Q_{\textrm{IR},a},Q_{\textrm{IR},b},Q_{\textrm{R}}) \notag  \\
  &=\sum_{\alpha} U_{\textrm{IR}}(Q_{\textrm{IR},\alpha}) + U_{\textrm{R}}(Q_{\textrm{R}}) \notag \\
  &-\frac{1}{2} g_{\textrm{IR}-\textrm{R}} (Q_{\textrm{IR},a}^2-Q_{\textrm{IR},b}^2)Q_{\textrm{R}} \\
  &-\frac{1}{2} h_{\textrm{IR}-\textrm{R}} (Q_{\textrm{IR},a}^2+Q_{\textrm{IR},b}^2) Q_{\textrm{R}}^2 \notag \\
  &-\frac{1}{2} \tilde{h}_{\textrm{IR}-\textrm{IR}}Q_{\textrm{IR},a}^2Q_{\textrm{IR},b}^2.\notag
\end{align}
Here, we consider the case where $Q_{\textrm{R}}$ belongs to the $B_{1g}$ representation.
Accordingly, the third-order coupling term takes the form $(Q_{\textrm{IR},a}^2-Q_{\textrm{IR},b}^2)Q_{\textrm{R}}$, while a term of the form $Q_{\textrm{IR},a}Q_{\textrm{IR},b}Q_{\textrm{R}}$ is not allowed.
Regarding fourth-order coupling terms, since $Q_{\textrm{R}}^2$ belongs to $A_{1g}$ representation, the same terms appear as the $E_u-A_{1g}$ coupling in Eq.~(\ref{eq:pot_Eu}).
When $Q_{\textrm{IR},a}^2=Q_{\textrm{IR},b}^2$, the third-order coupling term in Eq.~(\ref{eq:pot_B1g}) vanishes, and the minimum of the potential for $Q_{\textrm{R}}$ does not shift.
This condition is satisfied when the irradiation angle $\theta$ is set to $45^{\circ}$, $135^{\circ}$, $225^{\circ}$, and $315^{\circ}$.
As shown in Appendix~\ref{appendix_theta}, since the $E_u-A_{1g}$ coupling shows little dependence on the irradiation angle $\theta$, by choosing $\theta = 45^{\circ}, 135^{\circ}, 225^{\circ}$, or $315^{\circ}$, we assume that $\bar{h}$ can be enhanced without inducing a shift of the equilibrium position associated with the $B_{1g}$ modes.

\subsection{External field \label{subsec_field}}
Our theoretical proposal of the structural control can be experimentally realizable by the pump-probe method using a terahertz or mid-infrared laser~\cite{R.Mankowsky_2014,M.Forst_2011,M.Fechner_2024,M.Forst_2011,A.S.Disa_2020,T.G.Blank_2023}. 
The external driving force $F(t)$ considered in this study and the actual electric field $\bm{E}(t)$ are related by $F(t) = \bm{E}(t) \cdot \bm{Z}_{\alpha}$, where $\bm{Z}_{\alpha}$ is the effective charge of the phonon mode $\alpha$.
While first-principles evaluation of the effective charge for metals has been recently proposed by a dynamical extension of the effective charge~\cite{dynamicalZ1,dynamicalZ2,dynamicalZ3,dynamicalZ4}, it is still challenging to include the strong electron correlation effects into DFT calculations. Therefore, we do not evaluate $\bm{Z}_{\alpha}$ in this study, and hence we cannot show an explicit correspondence between the electric-field strength and $F_0$. Nevertheless, by assuming a typical value of $|\bm{Z}_{\alpha}| \sim 0.1\ e\,{\rm{u}}^{-1/2}$~\cite{A.S.Disa_2020}, the electric field strength is estimated to be $|\bm{E}| \sim 10$ MV cm$^{-1}$ for $F_0 = $ 10 meV $\mbox{\AA}^{-1} {\rm{u}}^{-1/2}$.

Although the amplitude of the external field mainly used in this study ($F_0 = 60\,\rm{meV} \mbox{\AA}^{-1} \rm{u}^{-1/2}$) may be significantly larger than experimentally accessible values, iron-based superconductors are highly sensitive to structural changes.
Therefore, even a small light-induced structural modification may result in experimentally detectable signatures of enhanced superconductivity under photoexcitation.

\section{Conclusion\label{sec_conclusion}}
In this study, we have investigated the possibility of controlling the crystal structure of the iron-based superconductor \ce{LaFeAsO} via light irradiation based on nonlinear phononics.
We have constructed the anharmonic lattice potential using first-principles calculations and calculated phonon dynamics by solving the equation of motion for the constructed potential.
We have found that the anion height $h$ increases when the in-plane vibrational IR mode of the Fe-centered tetrahedron is selectively excited.
We have also predicted the change in electronic properties from the equilibrium to the optically excited state by performing band calculations based on the time-averaged crystal structure.
Then we have found that the resulting modification of the band structure is associated with an increase in $h$, implying the possibility to enhance superconductivity.
We expect that our findings not only broaden the potential of photoinduced superconductivity in iron-based superconductors, but also stimulate theoretical and experimental studies on light-induced control of crystal structures based on nonlinear phononics.

\begin{acknowledgments}
We thank Takumi Fukuda, Takeshi Suzuki, Terumasa Tadano for fruitful discussions.
This work was supported by Grants-in-Aid for Scientific Research from JSPS, KAKENHI Grant No.~JP24K06939, No.~JP24H00191, No.~JP24K01333, and No.~JP25K08457.
S.K. and M.O. were supported by JST FOREST Program, Grant No. JPMJFR212P. 

\end{acknowledgments}

\appendix
\section{Eigenmodes for the $\Gamma$ point phonons and point group analysis\label{appendix_group}}
We present the frequencies and irreducible representations of all eigenmodes of LaFeAsO at the $\Gamma$ point in Table~\ref{table_eigenmodes}.
The phonon modes in LaFeAsO are classified according to $4/mmm$ point group symmetry, for which the product table is shown in Table~\ref{table_product}.
As described in the main text, the $A_{2u}$ and $E_u$ phonon modes have odd parity, whereas the $A_{1g}$, $B_{1g}$, and $E_g$ modes have even parity with respect to space inversion symmetry.
If the third-order coupling term of the form $Q_{\textrm{IR}}^2Q_{\textrm{R}}$ is included in the Hamiltonian, the overall symmetry of this term must belong to the $A_{1g}$ representation.
For the $A_{2u}$ modes, the direct product satisfies $A_{2u}\otimes A_{2u} = A_{1g}$, which implies that the coupled Raman mode $Q_{\textrm{R}}$ must belong to $A_{1g}$, i.e., $A_{2u}\otimes A_{2u} \otimes A_{1g} = A_{1g}$.
In contrast, for the $E_u$ case, the direct product is given by $E_u \otimes E_u = A_{1g}\oplus A_{2g}\oplus B_{1g}\oplus B_{2g}$.
The direct product of the term $Q_{\textrm{IR}}^2Q_{\textrm{R}}$ is
\begin{align*}
  E_u \otimes E_u \otimes A_{1g} &= A_{1g}\oplus A_{2g}\oplus B_{1g}\oplus B_{2g}\\
  E_u \otimes E_u \otimes B_{1g} &= A_{1g}\oplus A_{2g}\oplus B_{1g}\oplus B_{2g}\\
  E_u \otimes E_u \otimes E_{g}  &= 4E_g,
\end{align*}
for $A_{1g}$, $B_{1g}$, and $E_g$ Raman modes, respectively.
Therefore, in this case, $Q_{\textrm{R}}$ can belong to $A_{1g}$ and $B_{1g}$.

As a result, when the $A_{2u}$ mode is selected as IR mode, only the $A_{1g}$ mode couples as Raman mode.
On the other hand, when the $E_u$ mode is excited, the $A_{1g}$ and $B_{1g}$ modes are, in principle, allowed to couple.
Nevertheless, for the reasons explained in Sec.~\ref{sec_harmonic}, we focus exclusively on the $A_{1g}$ Raman mode in the present study.

\begin{table}[t]
\centering
  \begin{tabular}{ccc}
  \hline\hline
  label  &  Irr.      &$\omega/2\pi$ (THz)\\ \hline
  1, 2, 3 & acoustic & 0      \\
  4, 5   & $E_u$        & 2.043  \\
  6     & $A_{2u}$     & 2.682  \\
  7, 8   & $E_g$        & 3.535  \\
  9, 10  & $E_g$        & 4.302  \\
  11    & $A_{1g}$     & 5.562  \\
  12    & $A_{1g}$     & 6.240  \\
  13    & $B_{1g}$     & 6.650  \\
  14    & $A_{2u}$     & 7.839  \\
  15, 16 & $E_u$        & 8.294  \\
  17    & $B_{1g}$     & 8.654  \\
  18, 19 & $E_g$        & 8.677  \\
  20, 21 & $E_u$        & 9.297  \\
  22    & $A_{2u}$     & 12.308 \\
  23, 24 & $E_g$        & 13.142\\ \hline \hline
  \end{tabular}
  \caption{All phonon modes at the $\Gamma$ point in LaFeAsO.}
  \label{table_eigenmodes}
\end{table}

  \begin{table*}[ht]
    \centering
    \begin{tabular}{c|cccccccccc}
    \hline\hline
    $\times$ & $A_{1g}$ & $A_{2g}$ & $B_{1g}$ & $B_{2g}$ & $E_g$ & $A_{1u}$ & $A_{2u}$ & $B_{1u}$ & $B_{2u}$ & $E_u$ \\
    \hline
    $A_{1g}$ & $A_{1g}$ & $A_{2g}$ & $B_{1g}$ & $B_{2g}$ & $E_g$ & $A_{1u}$ & $A_{2u}$ & $B_{1u}$ & $B_{2u}$ & $E_u$ \\
    $A_{2g}$ & $\cdot$ & $A_{1g}$ & $B_{2g}$ & $B_{1g}$ & $E_g$ & $A_{2u}$ & $A_{1u}$ & $B_{2u}$ & $B_{1u}$ & $E_u$ \\
    $B_{1g}$ & $\cdot$ & $\cdot$ & $A_{1g}$ & $A_{2g}$ & $E_g$ & $B_{1u}$ & $B_{2u}$ & $A_{1u}$ & $A_{2u}$ & $E_u$ \\
    $B_{2g}$ & $\cdot$ & $\cdot$ & $\cdot$ & $A_{1g}$ & $E_g$ & $B_{2u}$ & $B_{1u}$ & $A_{2u}$ & $A_{1u}$ & $E_u$ \\
    $E_g$    & $\cdot$ & $\cdot$ & $\cdot$ & $\cdot$ &\begin{tabular}{l} $A_{1g}\oplus A_{2g}$\\$\oplus B_{1g}\oplus B_{2g}$\end{tabular} & $E_u$ & $E_u$ & $E_u$ & $E_u$ & \begin{tabular}{l} $A_{1u}\oplus A_{2u}$\\$\oplus B_{1u}\oplus B_{2u}$\end{tabular}\\
    $A_{1u}$ & $\cdot$ & $\cdot$ & $\cdot$ & $\cdot$ & $\cdot$ & $A_{1g}$ & $A_{2g}$ & $B_{1g}$ & $B_{2g}$ & $E_g$ \\
    $A_{2u}$ & $\cdot$ & $\cdot$ & $\cdot$ & $\cdot$ & $\cdot$ & $\cdot$ & $A_{1g}$ & $B_{2g}$ & $B_{1g}$ & $E_g$ \\
    $B_{1u}$ & $\cdot$ & $\cdot$ & $\cdot$ & $\cdot$ & $\cdot$ & $\cdot$ & $\cdot$ & $A_{1g}$ & $A_{2g}$ & $E_g$ \\
    $B_{2u}$ & $\cdot$ & $\cdot$ & $\cdot$ & $\cdot$ & $\cdot$ & $\cdot$ & $\cdot$ & $\cdot$ & $A_{1g}$ & $E_g$ \\
    $E_u$    & $\cdot$ & $\cdot$ & $\cdot$ & $\cdot$ & $\cdot$ & $\cdot$ & $\cdot$ & $\cdot$ & $\cdot$ & \begin{tabular}{l} $A_{1g}\oplus A_{2g}$\\$\oplus B_{1g}\oplus B_{2g}$\end{tabular} \\
    \hline\hline
    \end{tabular}
    \caption{Direct product table of irreducible representations for the point group $4/mmm$~\cite{group1,group2}. Note that the table is symmetric.}
    \label{table_product}
\end{table*}

\section{Effect of damping\label{appendix_damping}}
\begin{figure}[ht]
  \centering
  \includegraphics[width=0.85\linewidth]{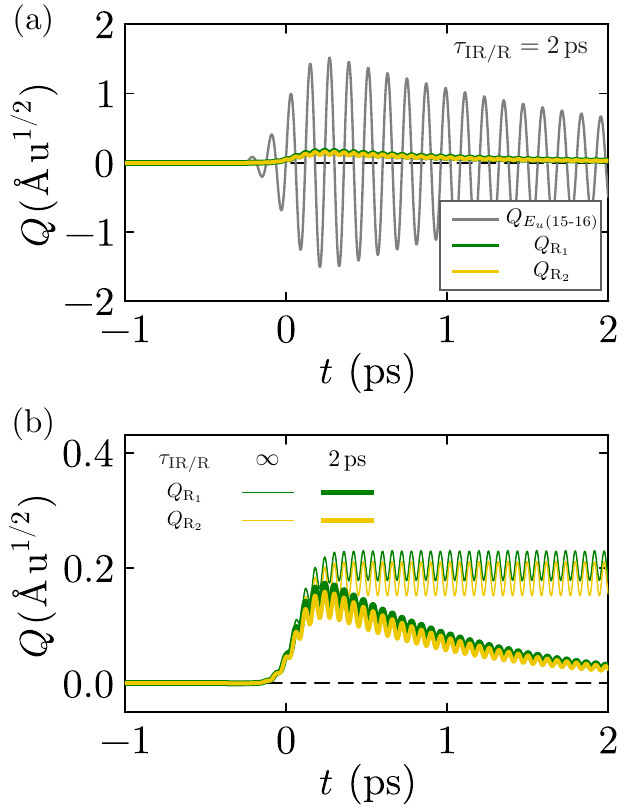}
  \caption{
    (a) Time evolution of the phonon normal coordinates of $E_{u}(15,16)$ and $A_{1g}$ modes including a damping term with $\tau_{\textrm{IR/R}}=2\,\textrm{ps}$. 
    (b) Comparison of the time evolution of Raman mode normal coordinates with and without a damping term. Thin and thick curves represent the cases with $\tau\to\infty$ ($\gamma=0$) and $\tau=2\,\textrm{ps}$, respectively.
  }
  \label{fig_damp}
\end{figure}
In this section, we present the results of phonon dynamics with finite phonon lifetimes.
The effect of the lifetime is taken into account by phenomenologically introducing a damping term into the equation of motion~\cite{R.Mankowsky_2014,Y.Zeng_2023,S_Kamiyama_2025}. 

The equations of motions for the phonon modes with the $E_u-A_{1g}$ coupling are
\begin{subequations}
\label{eq:eom_damp_Eu}
\begin{align}
  \ddot{Q}_{\textrm{IR},a} + 2\gamma_{\rm{IR}} \dot{Q}_{\textrm{IR},a} &= F(t)\cos\theta - \sum_{i}\frac{\partial  V_{E_u-A_{1g}}}{\partial Q_{\textrm{IR},a}}  \\
  \ddot{Q}_{\textrm{IR},b} + 2\gamma_{\rm{IR}} \dot{Q}_{\textrm{IR},b} &= F(t)\sin\theta - \sum_{i}\frac{\partial  V_{E_u-A_{1g}}}{\partial Q_{\textrm{IR},b}}  \\
  \ddot{Q}_{\textrm{R}_i} + 2\gamma_{\rm{R}_i} \dot{Q}_{\rm{R}_i} &= - \frac{\partial  V_{E_u-A_{1g}}}{\partial Q_{\textrm{R}_i}}\quad (i=1,2). 
\end{align}
\end{subequations}
$\gamma_{\rm{IR/R}}$ is the damping constant of the IR/Raman mode, which corresponds to the inverse of phonon lifetime $\tau_{\rm{IR/R}}$, i.e., $\gamma_{\rm{IR/R}} = 1/\tau_{\rm{IR/R}}$.
Here, we set $\tau_{\rm{IR/R}} = 2\,\rm{ps}$.
The results of the phonon dynamics for the $E_{u}(15,16)$ and $\textrm{R}_1,\textrm{R}_2$ modes are shown in Fig.~\ref{fig_damp}(a).
In Fig.~\ref{fig_damp}(b), the thin and thick lines indicate the time evolution of $Q_{\textrm{R}_1}$ and $Q_{\textrm{R}_2}$ without and with damping, respectively, where the thin lines correspond to Fig.~\ref{fig_dynamics_all}(c).
We confirm that the damped normal coordinates of $\textrm{R}_1$ and $\textrm{R}_2$ do not differ significantly from the undamped case for a while after light irradiation.

\section{Bond angle $\alpha$ \label{appendix_alpha}}

In this section, we investigate the change in $\alpha$ after light irradiation.
As mentioned in Sec.~\ref{sec_opt}, since the calculated crystal structures do not quantitatively match the experimental ones, we evaluate changes in $\alpha$ similar to that used for $h$.
First, we evaluate the values $\alpha_{\ce{La}}$ and $\alpha_{\ce{Sm}}$, bond angle of LaFeAsO and SmFeAsO, for both calculated and experimental structures, respectively.
Then we compare the difference $\Delta \alpha_{\ce{La}-\ce{Sm}} = \alpha_{\ce{Sm}}-\alpha_{\ce{La}}$ of calculations with that from experiments, as shown in Table~\ref{table_alpha}.

In Sec.~\ref{sec_dynamics}, we demonstrate that $\bar{h}$ increases the most when the $E_{u}(15,16)$ is optically excited.
We evaluate the value of $\alpha$ in time-averaged crystal structure after the $E_{u}(15,16)$ is stimulated with $F_0 = 60\,\rm{meV} \mbox{\AA}^{-1} \rm{u}^{-1/2}$.
The value is $\alpha=117.7^{\circ}$, decreasing around $1^{\circ}$ from its equilibrium value.
Therefore, we find that light irradiation also drives $\alpha$ toward the value of SmFeAsO.
\begin{table}[htbp]
  \centering
  \begin{tabular}{ccc}
    \hline\hline
    $\alpha$~(deg) & calc. & exp. \\ \hline
    $\alpha_{\ce{La}}$ & 118.7 & 113.6(1) \\
    $\alpha_{\ce{Sm}}$ & 116.3 & 110.5(1) \\
    $\Delta \alpha_{\ce{La}-\ce{Sm}}$ & 2.33 & 3.10(14) \\ \hline\hline
  \end{tabular}
  \caption{
    Bond angle $\alpha$ of LaFeAsO ($\alpha_{\ce{La}}$) and SmFeAsO ($\alpha_{\ce{Sm}}$) for both calculated and experimental structures.
  $\Delta \alpha_{\ce{La}-\ce{Sm}}$ is the difference in $\alpha$ between the two compounds, $\Delta \alpha_{\ce{La}-\ce{Sm}} = \alpha_{\ce{La}}-\alpha_{\ce{Sm}}$.
  The experimental values are taken from Refs.~\cite{Y_Kamihara_2008,A_Martinelli_2008}.}  
  \label{table_alpha}
\end{table}

\section{Dependence of the irradiation angle $\theta$ for the case of $E_u-A_{1g}$ coupling}\label{appendix_theta}
\begin{figure}[ht]
  \centering
  \includegraphics[width=1.0\linewidth]{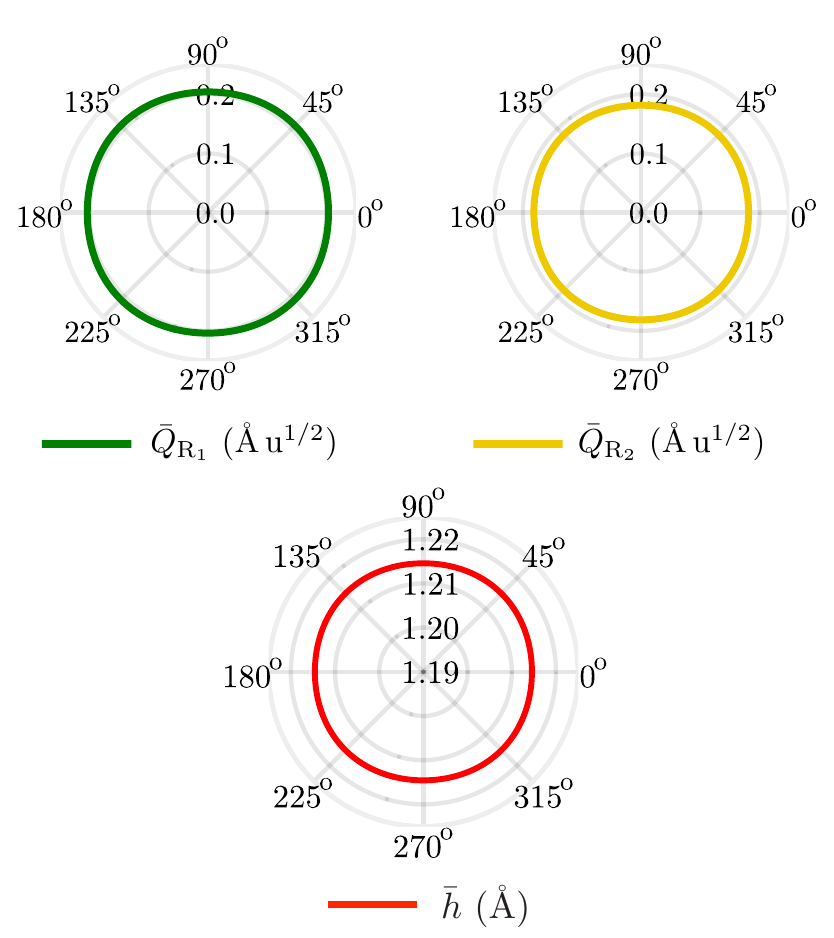}
  \caption{Irradiation angle $\theta$ dependence of $\bar{Q}_{\textrm{R}_1}$, $\bar{Q}_{\textrm{R}_2}$, and $\bar{h}$.
  }
  \label{fig_theta}
\end{figure}

In Fig.~\ref{fig_theta}, we present the dependence of the time-averaged values $\bar{Q}_{\textrm{R}}$ and $\bar{h}$ on the irradiation angle $\theta$, in accordance with Eq.~(\ref{eq:eqm_Eu}).
Here, we take the IR and Raman modes to be the $E_u(15,16)$ and the $A_{1g}$ modes ($\textrm{R}_1$, $\textrm{R}_2$), respectively.

Focusing on $\bar{Q}_{\textrm{R}}$, a slight enhancement is observed around $\theta = 45^{\circ}$, $135^{\circ}$, $225^{\circ}$, and $315^{\circ}$.
However, the variation is very small; for example, $\bar{Q}_{\textrm{R}_1}(0^{\circ}) = 0.203$~\AA~$\rm{u}^{1/2}$ and $\bar{Q}_{\textrm{R}_1}(45^{\circ}) = 0.208$~\AA~$\rm{u}^{1/2}$.
Overall, $\bar{Q}_{\textrm{R}}$ exhibits nearly isotropic behavior with respect to $\theta$, i.e., showing only weak angular dependence.
A similarly weak $\theta$ dependence is observed for $\bar{h}$.
Therefore, in the main text, we fix the irradiation angle to $\theta = 0^{\circ}$.

\bibliography{iron_paper}
\end{document}